\documentclass[aps,twocolumn,floatfix,amsmath,amssymb, showpacs, tightenlines]{revtex4}
\usepackage{amsfonts}
\usepackage{bbm}
\usepackage{amssymb}
\usepackage{graphicx}
\usepackage{epsfig,graphicx,times}
\usepackage{amstext}
\usepackage{amsmath}
\usepackage{latexsym}
\usepackage{bm}

\draft 

\begin{document}

\title{Generating Squeezed States of Nanomechanical Resonator}
\author{Wen Yi Huo$^{1}$ and Gui Lu Long$^{1,2,3}$ }

\address{
$^1$Key Laboratory for Atomic and Molecular NanoSciences and
Department of Physics,
Tsinghua University, Beijing 100084, China\\
$^2$  Tsinghua National Laboratory For Information Science and Technology, Beijing
100084, China\\
$^3$Institute of Microelectronics, Tsinghua University, Beijing 100084, China}

\date{\today}

\begin{abstract}
We propose a scheme for generating squeezed states in solid state circuits consisting
of a nanomechanical resonator (NMR),  a superconducting Cooper-pair box (CPB) and a
superconducting transmission line resonator (STLR). The nonlinear interaction between
the NMR and the STLR can be implemented by setting the external biased flux of the CPB
at  certain values. The interaction Hamiltonian between the NMR and the STLR is derived
by performing Fr$\rm\ddot o$hlich transformation on the total Hamiltonian of the
combined system. Just by adiabatically keeping the CPB at the ground state, we get the
standard parametric down-conversion Hamiltonian. The CPB plays the role of ``nonlinear
media", and the squeezed states of the NMR can be easily generated in a manner similar
to the three-wave mixing in quantum optics. This is the three-wave mixing in a
solid-state circuit.
\end{abstract}

\pacs{03.67.Lx,07.79.Cz, 33.40.+f} \maketitle

\section{introduction}

Mechanical harmonic oscillator plays an important role in the historical development of
quantum mechanics. The harmonic oscillator problem was one of the few completely
solvable problems when one began to learn quantum mechanics. Due to the macroscopic
nature, the experiments of mechanical harmonic oscillators didn't achieve much progress
for a quite long time, however. Recently, with the  development in quantum information
processing, people are now searching for  possible applications  of mechanical harmonic
oscillator  in this field, and evidence for quantized displacement in a nanomechanical
harmonic oscillator has been observed \cite{Gaidarzhy}. Many good physical ideas about
 the harmonic oscillator came up, though this system is rather simple. Squeezed
state was one of them. Squeezed state provides a good example of the interplay between
experiment and theory in the development of quantum mechanics. Statistical properties
of squeezed states have been widely investigated and the possibility of applying
squeezed states to study the fundamental quantum physics phenomena, as well as to
detect the gravitational radiation, has been recognized \cite{scully,bocko}.

Though the idea of squeezed states originated from mechanical harmonic oscillator, the
first experiment realization was the squeezed states of electromagnetic field in
nonlinear quantum optics\cite{slusher, wu}. In the nonlinear optical experiments,
three-wave and four-wave mixing were two main methods to  generate squeezed states. If
one injects low energy photons into a nonlinear optical medium, the second harmonic
generation may be induced and this forms a squeezed state, this procedure is called
three-wave mixing. Though the theory of four-wave mixing was more complicated, the
first squeezed state of electromagnetic field was implemented in this system. And
recently, the theory of generating squeezed states in a high-Q cavity was considered
\cite{Almeida}.

Recently, there has been significant progress in realizing quantum
optics in solid state electrical circuits. This new subject has a
nickname ``circuit quantum electrodynamics (circuit QED)"
\cite{blais, Chiorescu, geller1, huo07}. The extreme strong coupling
limit in cavity QED has been implemented experimentally with circuit
QED systems, such as superconducting charge qubit and
superconducting transmission line resonator (STLR) system
\cite{wallraff}, flux qubit and quantum $LC$ oscillator system
\cite{Chiorescu}, and phase qubit and nanomechanical resonator (NMR)
system ( which is a mechanical harmonic oscillator) \cite{geller2}.

Because of the extreme strong coupling in circuit QED, generating squeezed states in
such systems became very interesting and attracted much attention \cite{moon, zhou,
rabl, tian, Ruskov, xue}. In Ref. \cite{moon}, using two modes of the STLR coupled to a
superconducting charge qubit simultaneously, the authors studied microwave parametric
down-conversion and discussed the squeezed states of one mode of the STLR. In the other
Refs.\cite{zhou, rabl, tian, Ruskov}, the authors discussed the generation of squeezed
states of NMR. The generation of squeezed states of NMR in these schemes use either the
operations on the qubit or  dissipation and measurement to produce the needed
nonlinearity.

In this paper, we propose a scheme for generating squeezed states of NMR in circuit
QED, and it is similar to the three-wave mixing in optical cavity experiments. Our
proposal is based on the system consisting of a superconducting Cooper-pair box (CPB),
also a superconducting charge qubit, a NMR and a STLR. We find that with certain biased
conditions of the CPB, the nonlinear process can be implemented and the squeezed states
of NMR can easily be generated. The nonlinear interaction can be switched on and off at
will by changing the external biased flux of the CPB. By controlling the gate charge
$n_g>1/2$ and/or $n_g<0$, one can get different squeezed variables $x$ and/or $p$.
Compared with other schemes, our scheme is more simple and effective in generating the
squeezed states. In this scheme, the state of CPB is adiabatically kept in its ground
states, and does not  need qubit operations so that it avoids the restriction set by
the decoherence time of the qubit.

\section{The Model}

\begin{figure}[htbp]
\includegraphics[bb=130 563 368 740,width=8.3 cm]{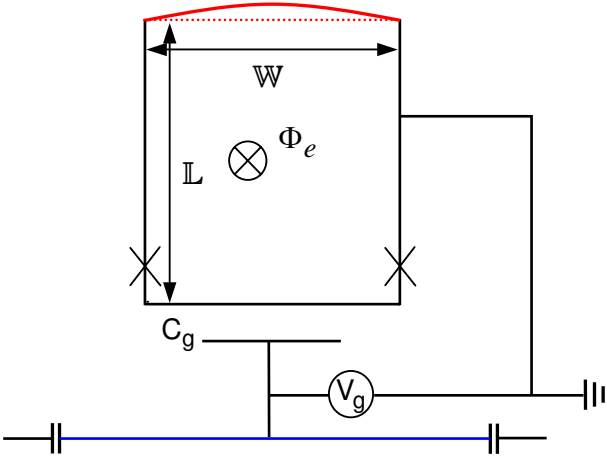}
\caption{(Color online) Schematic  diagram of combined system of a STLR, a NMR and a
superconducting CPB. The STLR (the blue line) is shown at the bottom, and at the top is
the NMR (the red line), which is one part of the SQUID.} \label{model}
\end{figure}

The system under consideration is shown in Fig.\ref{model}. A STLR (the blue line at
the bottom of Fig.\ref{model}) is placed close to a CPB. The state of CPB is separately
controlled by the gate voltage $V_g$ through a gate capacitance $C_g$. The CPB is also
coupled to a large superconductor reservoir through two identical Josephson junctions
with capacitance $C_J$ and Josephson energy $E_J$, and this forms a superconducting
quantum interference device (SQUID) and is also the basic configuration of
superconducting charge qubit. The SQUID configuration allows one to apply external flux
to control the Josephson energy. For superconducting charge qubits, the capacitance
$C_g$  is much less than $C_J$. In this regime a convenient basis is formed by the
charge states, characterized by the number of Cooper pairs $n$ on the CPB. In the
neighborhood of $n_g=1/2$, only two charge states $|N\rangle$ and $|N+1\rangle$ play a
role, while all other charge states, having a much higher energy, can be ignored. In
this case, the Hamiltonian of the CPB reads \cite{makhlin}
\begin{equation}
H_q=-\frac{1}{2}E_c(1-2n_g)\sigma_z-E_J\cos(\frac{\pi\Phi_e}{\Phi_0})\sigma_x,
\end{equation}
where $E_c=(2e)^2/(2C_{\Sigma})$ is the charging energy,
$C_{\Sigma}=2C_J+C_g$ is the total capacitance saw by the CPB,
$n_g=C_gV_g/(2e)$ is the gate charge induced by the gate voltage,
and $\Phi_e$ is the externally applied flux.

For a CPB fabricated inside a STLR, there is not only the dc voltage $V_g$ but also a
quantum part $V_q$ applied on the gate capacitance. The quantized voltage at the
antinode $z=L/(2k)$ of the STLR takes it's maximum amplitude \cite{blais}
\begin{eqnarray}\nonumber
V_q=\sum_kV_0^k(a_k+a_k^{\dagger}),&&
V_0^k=\sqrt{\frac{\hbar\omega_k}{Lc}}.
\end{eqnarray}
Here, $\omega_k=k\pi/(L\sqrt{lc})$, with $L$, $l$ and $c$ being the
length, the inductance and capacitance per unit length of the STLR,
respectively. At low temperatures, there is only one mode of the
STLR, say $\omega_k=\omega_a$, that couples to the CPB, then the
quantum voltage applied on the gate capacitance beomes
$V_q=V_0(a+a^{\dagger})$. The Hamiltonian of the joint system (STLR
and CPB) has a spin-boson form
\begin{eqnarray}\nonumber
H_1&=&-\frac{1}{2}E_c(1-2n_g)\sigma_z-E_J\cos(\frac{\pi\Phi_e}{\Phi_0})\sigma_x\\
&&+\frac{eC_gV_0}{C_{\Sigma}}(a+a^{\dagger})\sigma_z+\hbar\omega_aa^{\dagger}a.
\end{eqnarray}

Now we consider the coupling between the NMR and the CPB. A NMR is fabricated as a part
of the SQUID, which has a length $\mathbbm{L}$ and width $\mathbbm{W}$, as shown in
Fig.\ref{model}. Considering the small displacement of the NMR, the effective area of
the SQUID is $S=\mathbbm{W}(\mathbbm{L}+x)$, where $x$ is the displacement operator and
$x=\sqrt{\hbar/(2M\omega_b)}(b+b^{\dagger})$, with $\omega_b$ and $M$ being the
frequency and mass of the NMR, respectively. Thus the flux bias of the SQUID is
$\Phi_e=\Phi_e^0+B\mathbbm{W}x$, where $\Phi_e^0$ is the flux bias corresponding to the
equilibrium position of the NMR \cite{zhou}. With the total flux bias, the effective
Josephson coupling energy of the SQUID becomes
\begin{equation}
E_J(x)=-E_J\cos\frac{\pi(\Phi_e^0+B\mathbbm{W}x)}{\Phi_0}.
\label{potential}
\end{equation}
The Hamiltonian of the total system, including STLR, NMR and CPB
reads
\begin{eqnarray}\nonumber
H_2&=&-\frac{1}{2}E_c(1-2n_g)\sigma_z-E_J\cos\frac{\pi(\Phi_e^0+B\mathbbm{W}x)}{\Phi_0}\sigma_x\\
&&+\frac{eC_gV_0}{C_{\Sigma}}(a+a^{\dagger})\sigma_z+\hbar\omega_aa^{\dagger}a+\hbar\omega_bb^{\dagger}b.
\label{htotal}
\end{eqnarray} It's obviously that the Josephson
coupling energy Eq.(\ref{potential}) is a nonlinear function of $x$
and can be expanded to
\begin{eqnarray}\nonumber
E_J(x)&=&-E_J[\cos(\frac{\pi\Phi_e^0}{\Phi_0})\cos(\frac{\pi
B\mathbbm{W}x}{\Phi_0})\\
&&-\sin(\frac{\pi\Phi_e^0}{\Phi_0})\sin(\frac{\pi
B\mathbbm{W}x}{\Phi_0})]. \label{potential2}
\end{eqnarray}
In general, $x$ is very small, then the first and second term of Eq.(\ref{potential2})
can be discarded and expanded to the first order of $x$, respectively. In this way, one
can get the linearized function of Josephson coupling energy of $x$, and there is only
linear interaction terms remained in Hamiltonian (\ref{htotal}). As pointed by Zhou
{\it et al.} \cite{zhou}, the nonlinearity of the Josephson coupling energy in $x$
shouldn't be neglected all the time, and it would be important for the generation of
squeezed states.

\section{Squeezing of NMR}

In the past years, two cavities interacting with a two-level atom has been widely
studied \cite{benivegna, xie, sun},  they focused on the linear interaction in such
systems. Here, we study the nonlinear interaction in the STLR, NMR and CPB system. An
important nonlinear effect is the generation of squeezed states, which is an
outstanding task in quantum mechanics. In the following we show how we realize the
nonlinear interaction and how we can use the nonlinear interaction to generate squeezed
states. Our method is similar to the three-wave mixing in quantum optics.

In order to realize nonlinear interaction, one can bias the SQUID at
$\Phi_e^0=m\Phi_0$, here $m$ is an integer, and expand the Josephson energy to the
second order in $x$, then the Hamiltonian (\ref{htotal}) becomes
\begin{eqnarray}\label{hamiltonian}\nonumber
H_t&=&\hbar\omega_aa^{\dagger}a+\hbar\omega_bb^{\dagger}b-
\frac{1}{2}E_c(1-2n_g)\sigma_z-(-1)^mE_J\sigma_x\\
&&+\hbar\lambda_a(a^{\dagger}+a)\sigma_z+\hbar\lambda_b(b^{\dagger}+b)^2\sigma_x,
\end{eqnarray}
where
\begin{eqnarray}\nonumber
\lambda_a=\frac{eC_g}{\hbar
C_{\Sigma}}\sqrt{\frac{\hbar\omega_a}{Lc}}, &
\lambda_b=\frac{(-1)^{m}E_J(\pi
B\mathbbm{W})^2}{4M\omega_b\Phi_0^2}.
\end{eqnarray}

In the following we choose the eigenenergy basis (spanned by
$|0\rangle=\sin(\theta/2)|N+1\rangle+\cos(\theta/2)|N\rangle$ and
$|1\rangle=\cos(\theta/2)|N+1\rangle-\sin(\theta/2)|N\rangle$) to
simplify the above Hamiltonian. Here,
$\theta=\tan^{-1}[(-1)^m2E_J/(E_c(1-2n_g))]$ is the mixing angle. In
the representation of eigenenergy basis of the CPB and  under the
rotating-wave approximation, the Hamiltonian (\ref{hamiltonian}) is
simplified to
\begin{eqnarray}\label{hra}\nonumber
H&=&\hbar\omega_aa^{\dagger}a+\hbar\omega_bb^{\dagger}b-
\frac{1}{2}\hbar\Omega\rho_z\\
&&+\hbar g_a(a^{\dagger}\rho_-+a\rho_+)+\hbar
g_b(b^{\dagger2}\rho_-+b^2\rho_+),
\end{eqnarray}
where $\Omega=\sqrt{E_c^2(1-2n_g)^2+4E_J^2}/\hbar$,
$\rho_z=|0\rangle\langle0|-|1\rangle\langle1|$,
$\rho_+={\rho_-}^{\dagger}=|1\rangle\langle0|$,
$g_a=-\lambda_a\sin\theta$ and $g_b=\lambda_b\cos\theta$.

Assuming the detunings between the CPB and the STLR and the NMR satisfy the large
detuning limits, that is $|\Delta_a|=|\Omega-\omega_a|,\hspace{4pt}
|\Delta_b|=|\Omega-2\omega_b|\gg g_a,\hspace{4pt}g_b$,
then the variables of the CPB can be adiabatically eliminated by performing the
Fr$\rm\ddot{o}$hlich transformation \cite{frohlich} on the total Hamiltonian
(\ref{hra}). Dividing the Hamiltonian (\ref{hra}) into two parts $H=H_0+H_I$, with
\begin{eqnarray}
H_0=\hbar\omega_aa^{\dagger}a+\hbar\omega_bb^{\dagger}b-
\frac{1}{2}\hbar\Omega\rho_z,\nonumber
\end{eqnarray}
and
\begin{eqnarray}
H_I=\hbar g_a(a^{\dagger}\rho_-+a\rho_+)+\hbar
g_b(b^{\dagger2}\rho_-+b^2\rho_+).\nonumber
\end{eqnarray}
Apply a unitary transformation $H_S=e^{-S}He^{S}$ with the generator
$S=g_a(a^{\dagger}\rho_--a\rho_+)/\Delta_a+g_b(b^{\dagger2}\rho_--b^2\rho_+)/\Delta_b$,
and expand $H_S$ to second order in $g_i/\Delta_i$ ($i=a, b$), we obtain the effective
Hamiltonian
\begin{eqnarray}\nonumber
H_S&=&\hbar(\omega_a-\frac{g_a^2}{\Delta_a}\rho_z)a^{\dagger}a
-\frac{1}{2}\hbar(\Omega+\frac{2g_b^2}{\Delta_b}+\frac{g_a^2}{\Delta_a})\rho_z\\
\nonumber
&&+\hbar(\omega_b+\frac{2g_b^2}{\Delta_b}-\frac{g_b^2}{\Delta_b}\rho_z
-\frac{g_b^2}{\Delta_b}b^{\dagger}b\rho_z)b^{\dagger}b\\
&&-\frac{1}{2}\hbar
g_ag_b(\frac{1}{\Delta_a}+\frac{1}{\Delta_b})(a^{\dagger}b^2+b^{\dagger2}a)\rho_z.
\end{eqnarray}
If the CPB is adiabatically kept on the ground state, the effective Hamiltonian becomes
\begin{eqnarray}\nonumber
H_S&=&\hbar(\omega_a-\frac{g_a^2}{\Delta_a})a^{\dagger}a+\hbar(\omega_b+\frac{g_b^2}{\Delta_b}-\frac{g_b^2}{\Delta_b}b^{\dagger}b)b^{\dagger}b
\\
&&-\frac{1}{2}\hbar
g_ag_b(\frac{1}{\Delta_a}+\frac{1}{\Delta_b})(b^{\dagger2}a+a^{\dagger}b^2).
\end{eqnarray}
In general, $\omega_a\gg g_a^2/\Delta_a$ and $\omega_b\gg
g_b^2/\Delta_b$, then the effective Hamiltonian can be approximated
as
\begin{equation}\label{pdc}
H_S=\hbar\omega_aa^{\dagger}a+\hbar\omega_bb^{\dagger}b
-\frac{1}{2}\hbar
g_ag_b(\frac{1}{\Delta_a}+\frac{1}{\Delta_b})(b^{\dagger2}a+a^{\dagger}b^2).
\end{equation}
In the interaction picture, the Hamiltonian (\ref{pdc}) reads
\begin{equation}
H_I(t)=-\frac{1}{2}\hbar
g_ag_b(\frac{1}{\Delta_a}+\frac{1}{\Delta_b})(b^{\dagger2}ae^{i\delta
t}+a^{\dagger}b^2e^{-i\delta t}),
\end{equation}
where $\delta=2\omega_b-\omega_a$. If $\delta=0$, that is
$\omega_a=2\omega_b$ and $\Delta_a=\Delta_b=\Delta$, in this
situation the Hamiltonian becomes
\begin{equation}\label{hinter}
H_I(t)=\hbar\kappa(b^{\dagger2}a+a^{\dagger}b^2),
\end{equation}
where $\kappa=-g_ag_b/\Delta$ is the coupling constant between the
STLR and NMR.

In the parametric approximation, the Hamiltonian (\ref{hinter})
becomes
\begin{equation}
H_I(t)=\hbar\kappa\beta(b^{\dagger2}e^{-i\phi}+b^2e^{i\phi}),
\end{equation}
where $\beta$ and $\phi$ is the amplitude and phase of the STLR which is in a coherent
state. The time evolution operator of the NMR in the interaction picture reads
\begin{equation}\label{top}
U(t)=e^{-i\kappa\beta t(b^{\dagger2}e^{-i\phi}+b^2e^{i\phi})}.
\end{equation}
In fact, the time evolution operator (\ref{top}) is the squeezed
operator of the NMR. For a time duration $\tau$, the squeezed
operator reads
\begin{equation}
S(\xi)=e^{-i\frac{\xi}{2}(b^{\dagger2}e^{-i\phi}+b^2e^{i\phi})},
\end{equation}
where $\xi=2\kappa\beta\tau$ is the effective squeezed parameter.
For the NMR initially in the vacuum state $|0\rangle$ and
$\phi=\pi/2$, the variance in the two quadratures
$x=x_0(b^{\dagger}+b)$ and $p=ip_0(b^{\dagger}-b)$, where
$x_0=\sqrt{\hbar/(2M\omega_b)}$ and $p_0=\sqrt{\hbar M\omega_b/2}$,
can be calculated directly using the transformation
$S^{\dagger}(\xi)bS(\xi)=b\cosh\xi-b^{\dagger}\sinh\xi$,
\begin{subequations}
\begin{eqnarray}
&&\Delta x=\sqrt{\langle x^2\rangle-(\langle
x\rangle)^2}=x_0e^{-\xi},\\
&&\Delta p=\sqrt{\langle p^2\rangle-(\langle
p\rangle)^2}=p_0e^{\xi}.
\end{eqnarray}
\end{subequations}
And the NMR is in the squeezed state
\begin{equation}
|\xi\rangle=S(\xi)|0\rangle=e^{-\frac{\xi}{2}(b^{\dagger2}-b^2})|0\rangle.
\end{equation}
If the NMR in the coherent state $|\alpha\rangle$ initially, one can
generate the ideal squeezed state
$|\xi,\alpha\rangle=S(\xi)|\alpha\rangle$.

To estimate the squeezed efficiency, we choose the typical parameters in current
solid-state circuits experiments as follows: $E_J/2\pi=4$ GHz, $\Omega/2\pi=10$ GHz,
$\omega_a/2\pi=2\omega_b/2\pi=3$ GHz, $C_g/C_{\Sigma}=0.1$, $B=0.2$ T, $\mathbb{W}=1
\mu$m, $V_0=2 \mu$V, and $x_0=10^{-12}$ m. We can get two different values of coupling
constant $\kappa/2\pi\approx0.6$ Hz with $n_g>1/2$ and $\kappa/2\pi\approx-0.6$ Hz with
$n_g<1/2$.

In the above discussion, we have assumed an ideal situation in which the noise and
fluctuations were not included. The noise in the CPB may not be considered in our
scheme because the CPB can  naturally stay in it's ground state at low temperatures for
a very long time. Here we investigate the effect of a finite linewidth $D$ induced by
phase fluctuations in driving coherent state on the squeezing properties of the NMR, as
has been done in three-wave mixing in quantum optics.

\begin{figure}[htbp]
\includegraphics[bb=90 498 503 772,width=8.3 cm]{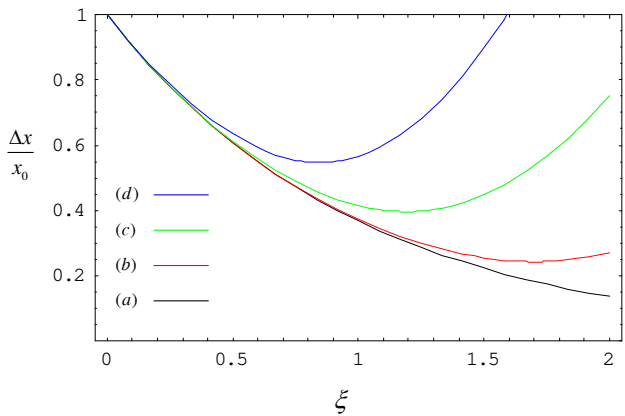}
\caption{(Color online) Squeezed efficiency $\Delta x/x_0$ versus $\xi$ for different
ratios (a) $D/2\kappa\beta=0$, (b) $D/2\kappa\beta=0.001$, (c) $D/2\kappa\beta=0.01$,
(d) $D/2\kappa\beta=0.05$.} \label{noise}
\end{figure}

For the linewidth $D$, and in the limit $D\ll
\tau^{-1}\ll2\kappa\beta$, the variances in the two quadratures are
given by \cite{scully}
\begin{subequations}
\begin{eqnarray}\label{sq}
&&\Delta x=x_0\sqrt{e^{-2\xi}+(\frac{D\tau}{2})e^{2\xi}},\\
&&\Delta p=p_0e^{\xi}\sqrt{1-2D\tau}.
\end{eqnarray}
\end{subequations}
In Fig.\ref{noise}, we have plotted $\Delta x/x_0$ versus $\xi$ for
various ratios of $D/2\kappa\beta$. The variances in the amplitude
of $x$ increases due to the phase fluctuations in the driving
coherent state. There is a minimum in $\Delta x/x_0$ which decreases
with increasing $D/2\kappa\beta$. It's evident from Eq.(\ref{sq})
that the phase fluctuation in driving coherent state would affect
the squeezed efficient severely with increasing time $\tau$ due to
the existence of $e^{2\xi}$.

\section{discussion and conclusion}

Our scheme is much different from the existing ones  such as
Refs.\cite{moon, zhou, rabl, tian, Ruskov}. In our scheme, it
doesn't need any operations on the CPB and doesn't use dissipation
and measurement to generate the needed nonlinearity. It just needs
to adiabatically keep the CPB in it's ground state. Our scheme can
greatly decrease the effect of the decoherence of the CPB on the
squeezed efficency. The CPB plays the role of nonlinear media as in
quantum optics, and our result is similar to the three-wave mixing.
By controlling $n_g>1/2$ and/or $N_g<1/2$, we can get different
squeezed variables $x$ and/or $p$.

In conclusion, we proposed a scheme for generating squeezed states
in solid state circuits system. In such a system, a NMR is
fabricated as a part of a SQUID, which consists of a CPB, and a STLR
is capacitively coupled to the CPB. The nonlinear interaction
between the CPB and the NMR can be implemented by setting the
external biased flux of the CPB at some certain values. By
performing Fr\"{o}hlich transformation, we can get the nonlinear
Hamiltonian of parametric down-conversion of the STLR-NMR system. In
our scheme, the CPB plays the role of ``nonlinear media" and the
squeezed states of the NMR can be generated in a manner similar to
the three-wave mixing in quantum optics.

\begin{acknowledgments}
This work is supported by the National Fundamental Research Program
Grant No. 2006CB921106, China National Natural Science Foundation
Grant Nos. 10325521, 60433050, 60635040,  the SRFDP program of
Education Ministry of China, No. 20060003048 and the Key grant
Project of Chinese Ministry of Education No.306020.
\end{acknowledgments}


\begin{thebibliography}{99}

\bibitem{Gaidarzhy} A. Gaidarzhy, G. Zolfagharkhani, R. L. Badzey and P.
Mohanty, Phys. Rev. Lett. {\bf 94}, 030402 (2005);
\bibitem{scully} M. O. Scully and M. S. Zubairy,
{\it Quantum Optics}, Cambridge University Press, Cambridge, 1997.
\bibitem{bocko}M. F.
Bocko and R. Onofrio, Rev. Mod. Phys. {\bf 68}, 755 (1996).
\bibitem{slusher} R. E. Slusher, L. W. Hollberg, B. Yurke, J. C. Mertz, and J. F.
Valley, Phys. Rev. Lett. {\bf 55}, 2409 (1985).
\bibitem{wu} L. A. Wu, H. J. Kimble, J. L. Hall and H. Wu, Phys.
Rev. Lett. {bf 57}, 2520 (1987).
\bibitem{Almeida} N. G. de Almeida, R. M. Serra, C. J. Villas-B\"{o}as and M. H. Y.
Moussa, Phys. Rev. A {\bf 69}, 035802 (2004).
\bibitem{blais} A. Blais, R.-S. Huang, A. Wallraff, S. M. Girvin, and R. J.
Schoelkopf, Phys. Rev. A {\bf 69}, 062320 (2004).
\bibitem{Chiorescu} I. Chiorescu, P. Bertet, K. Semba, Y. Nakamura,
C. J. P. M. Harmans and J. E. Mooij, Nature {\bf 431}, 159 (2004).
\bibitem{geller1} A. N. Cleland and M. R. Geller, Phys. Rev. Lett. {\bf 93},
070501 (2004).
\bibitem{huo07} W. Y. Huo and G. L. Long, arXiv:quant-ph/0702104.
\bibitem{wallraff} A. Wallraff, D. I. Schuster, A. Blais, L. Frunzio, R.- S. Huang,
J. Majer, S. Kumar, S. M. Girvin, and R. J. Schoelkopf, Nature {\bf
431}, 162 (2004).
\bibitem{geller2} M. R. Geller and A. N. Cleland, Phys. Rev. A {\bf
71}, 032311 (2005).
\bibitem{moon} K. Moon and S. M. Girvin, Phys. Rev. Lett. {\bf95},
140504 (2005).
\bibitem{zhou} X. X. Zhou and A. Mizel, Phys. Rev. Lett.{\bf 97},
267201 (2006).
\bibitem{rabl} P. Rabl, A. Shnirman and P. Zoller, Phys. Rev. B {\bf
70}, 205304 (2004).
\bibitem{tian} L. Tian and R. W. Simmonds, arXiv: cond-mat/0606787.
\bibitem{Ruskov}R. Ruskov, K. Schwab and A. N. Korotkov, Phys. Rev. B
{\bf 71}, 235407 (2005).
\bibitem{xue} F. Xue, Y. X. Liu, C. P. Sun and F. Nori, arXiv:
quant-ph/0701209.
\bibitem{benivegna} G. Benivegna and A. Messina, J. Mod. Opt. {\bf 41}, 907 (1994).
\bibitem{xie} Y. B. Xie, J. Mod. Opt. {\bf 42}, 2239 (1994).
\bibitem{sun} C. P. Sun, L. F. Wei, Y. X. Liu, and F. Nori,
Phys. Rev. A {\bf 73}, 022318 (2006).
\bibitem{makhlin} Y. Makhlin, G. Sch\"{o}n and A. Shnirman,
Rev. Mod. Phys. {\bf 73}, 357 (2001).
\bibitem{armour} A. D. Armour, M. P. Blencowe and K. C. Schwab,
Phys. Rev. Lett. {\bf 88}, 148301 (2002).
\bibitem{irish} E. K. Irish and K.
Schwab, Phys. Rev. B {\bf 68}, 155311 (2003).
\bibitem{martin} I. Martin, A. Shnirman, L. Tian and
P. Zoller, Phys. Rev. B {\bf 69},125339 (2004).
\bibitem{frohlich} H. Fr$\rm\ddot{o}$hlich, Phys. Rev. {\bf 79}, 845 (1950).

\end{thebibliography}
\end{document}